# Integrated near-field thermo-photovoltaics for on-demand heat recycling

Gaurang R. Bhatt[1], Bo Zhao[2], Samantha Roberts[1], Ipshita Datta[1], Aseema Mohanty[1], Tong Lin[1], Jean-Michel Hartmann[3], Raphael St-Gelais[4], Shanhui Fan[2], Michal Lipson[1*]

[1] Department of Electrical Engineering, Columbia University, New York, New York 10027, USA
[2] Department of Electrical Engineering, Ginzton Laboratory, Stanford University, Stanford, California 94305, USA
[3] CEA · Laboratoire d'Électronique des Technologies de l'Information (LETI) Minatec Campus, Grenoble, France.
[4] Department of Mechanical Engineering, University of Ottawa, Ottawa, Ontario, K1N 6N5, Canada

*ml3745@columbia.edu

**The energy transferred via thermal radiation between two surfaces separated by nanometers distances (near-field) can be much larger than the blackbody limit [1–11]. However, realizing a reconfigurable platform that utilizes this energy exchange mechanism to generate electricity in industrial and space applications on-demand, remains a challenge [12–23]. The challenge lies in designing a platform that can separate two surfaces by a small and tunable gap while simultaneously maintaining a large temperature differential. Here, we present a fully integrated, reconfigurable and scalable platform operating in near-field regime that performs controlled heat extraction and energy recycling. Our platform relies on an integrated nano-electromechanical system (NEMS) that enables precise positioning of a large area thermal emitter within nanometers distances from a room-temperature germanium photodetector to form a thermo-photovoltaic (TPV) cell. We show over an order of magnitude higher power generation ($P_{gen} \sim 1.25\ \mu W \cdot cm^{-2}$) from our TPV cell by tuning the gap between a hot emitter ($T_E \sim 880$ K) and the cold photodetector ($T_D \sim 300$ K) from $\sim 500$ nm to $\sim 100$ nm. The significant enhancement in $P_{gen}$ at such small distances is a clear indication of near-field heat transfer effect. Our electrostatically controlled NEMS switch consumes negligible tuning power ($P_{gen}/P_{NEMS} \sim 10^4$) and relies on conventional silicon-based process technologies.**

Harvesting of thermal energy can be performed by relying on radiation from systems requiring dynamic temperature profiles through active tuning of distance between a cold photodetector placed at nanometers distance from a hot surface [23–27]. Strong heat exchange occurs at nanometer (near-field) distances due to evanescent modes, i.e., radiation modes that cannot be emitted from the hot body into the far field, but can evanescently couple from the hot to the cold surface when the separation is sub-wavelength. This radiative heat exchange is much stronger than blackbody limit at small gaps (d) between the hot and cold surfaces and reduces as $1/d^\alpha$ as the gap increases ($1 \leq \alpha \leq 2$; $\alpha$ is a geometry dependent factor) [24,28]. Electricity generation from heat exchangers and thermal management systems in industrial and space applications can be performed on-demand using this effect through precise control of the distance (d) [29,30]. The underlying theory of heat exchange via near-field radiation has been widely studied over the past decade and only recently demonstrated for energy generation using a lab-scale table-top experiment based on precision nano-positioning systems as well as in non-reconfigurable passive platforms [31,32].

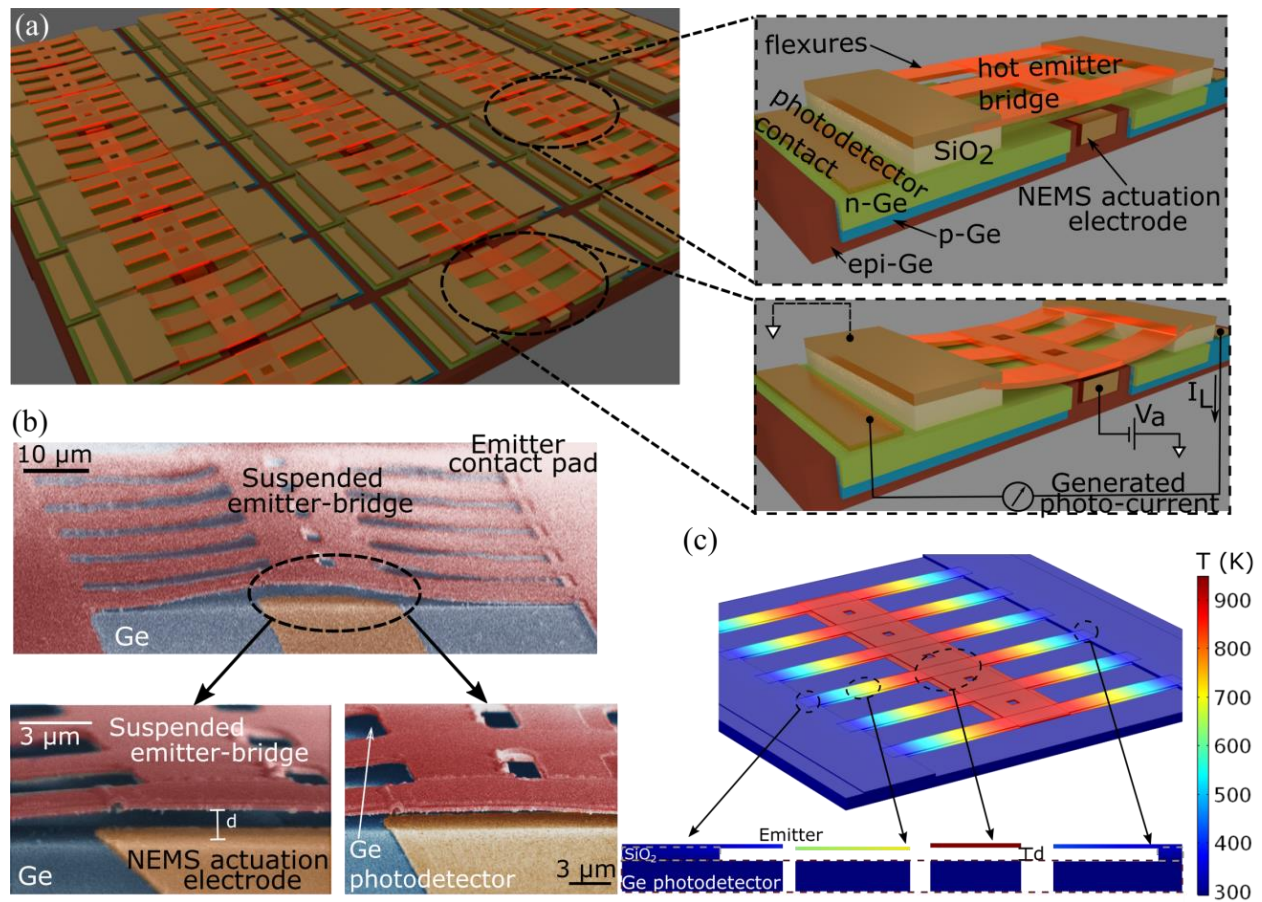

**Figure 1. (a)** Schematic of an array of thermo-photovoltaic cells demonstrated here. The platform consists of a nano-electromechanical system (NEMS) integrated with a photodetector cell. The NEMS consists of a suspended bridge ($80 \times 15$ μm$^2$) made of amorphous silicon (a-Si) anchored on two silicon dioxide (SiO$_2$) pads. The zoomed in schematic shows a single unit cell, where one can see the suspended bridge, the partially underlying actuation electrode and the photodetector region. The silicon substrate over which the cell is fabricated is not shown here. The central region of suspended bridge is brought closer to the underlying photodetector cell through application of actuation potential. The suspended bridge serves as mechanical support for a thin-film metallic thermal emitter, which can be heated to high temperature. The generated electricity from near-field radiative heat is collected at the photodetector terminals. **(b)** Scanning electron micrograph of the fabricated structures showing the emitter suspended over the Ge photodetector and NEMS actuation electrode. The holes in the suspended bridge facilitate etching of the sacrificial oxide layer during the release process. **(c)** Simulated temperature distribution in the structure when the emitter is heated up to high temperatures. The insets show the cross section of the structure at different positions. One can see that the when the emitter is hot (shown by red) the clamped edges at the pad and the substrate remain at room temperature (shown in blue), due to high thermal resistance of the thin-film emitter.

We demonstrate a scalable near-field TPV switch that allows us to control the heat flux through precise tuning of the distance between a hot suspended metallic thermal emitter and a room-temperature photodetector. The scalable platform is shown in Fig. 1a while the electron microscope image of an individual TPV cell is shown in Fig. 1b. The TPV cell consists of a suspended emitter, underlying photodetector and a pair of actuators for mechanical control. The suspended thermal emitter (80 × 15 μm²) consists of a metallic thin-film supported by a layer of amorphous silicon (a-Si) and anchored to silicon dioxide (SiO₂) pads with the help of multiple flexures (pad area: $A_{pad} \approx 450 \times 450$ μm², $t_{SiO_2} = 300$ nm). The suspended emitter overlaps the actuation electrode at each end ($A_{overlap} \approx 5 \times 15$ μm²) while the rest of the bridge area overlaps the active region of a photodetector. The in-plane photodetector and the actuation electrode are separated by a distance of 4 μm. The suspended emitter is brought closer to the underlying photodetector by virtue of electrostatic attraction through the application of an actuation potential $V_a$ to the actuators. The multiple flexures holding the central bridge of the emitter provide the required spring-like restoring force for controlled tuning of the gap. The TPV shown here relies on thin-film metallic emitter with high thermal resistance to achieve effective thermal insulation between the hot emitter and the underlying photodetector, allowing heat transfer only through radiation. Figure 1c shows the heat map of the structure when the emitter is heated to high temperatures. One can see that while the emitter temperature is over 900 K, the temperature of the photodetector embedded in the substrate remains at 300 K. Maintaining such a large temperature difference between the emitter and the photodetector is essential for energy harvesting purposes. According to the Carnot efficiency limit $\eta_{Carnot} = (T_H - T_C)/T_H$ where, $T_H$ and $T_C$ are temperatures of hot and cold surfaces respectively, the efficiency of a heat engine increases with a larger temperature difference between hot and cold sides. Furthermore, higher temperature differential also results in higher power density at the photodetector. The thermal circuit of the structure is provided in supplementary section 5. The details of the device dimensions are provided in supplementary section 1, and the fabrication procedure is provided in the methods section.

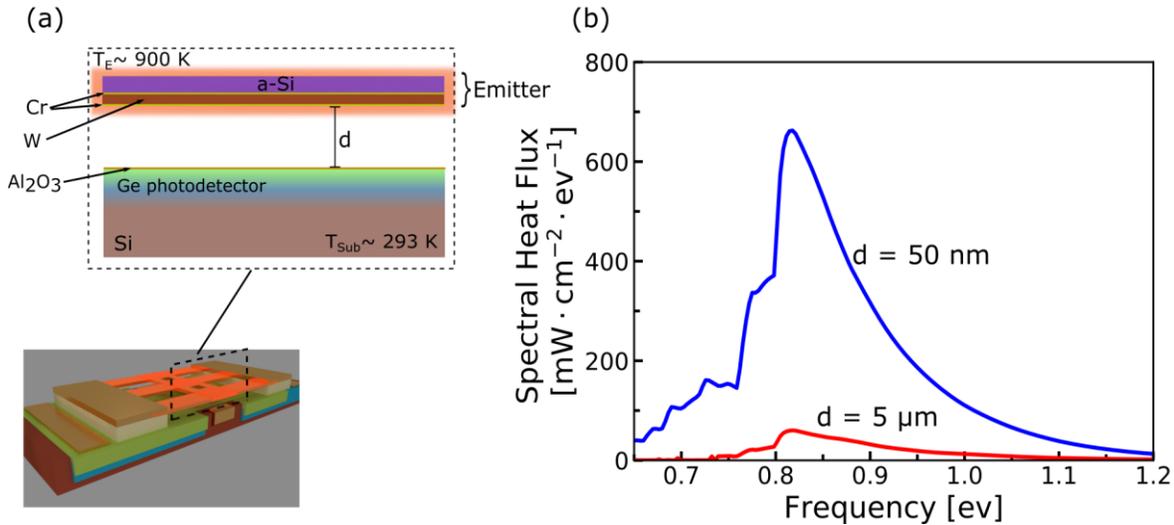

**Figure 2. Computational analysis of the heat transfer (a)** Schematic of the material stack of emitter and photodetector of the TPV. In the hot-stack, chromium (Cr) film helps in adhesion and nucleation of tungsten thin-film, while a-Si provides mechanical support when the bridge is suspended. The cold-stack is composed of Ge-on-silicon and covered with 10 nm alumina (Al₂O₃) to protect the surface and avoid any unwanted current leakage from bridge to photodetector in case of shorting during measurement. **(b)** Calculated spectral heat flux of thermal radiation absorbed by room-temperature germanium when the hot-emitter (aSi-Cr-W-Cr) is placed at near-field distances (50 nm, blue-line) and far-away (5 μm, red-line). The temperature of the emitter is assumed to be 900 K.

We choose a tungsten-chrome thin-film thermal emitter and a germanium photodetector to demonstrate electricity generation from near-field radiative heat transfer. Figure 2a shows the cross-section of our TPV cell with various films stacks that constitute the emitter and the photodetector. We choose tungsten (W) as thermal emitter due to its thermal stability at high temperatures and compatibility with silicon processing technology. Thin-film chromium is required to achieve adhesion and nucleation of W films to the silicon bridge on the top and to the underlying sacrificial layer during the fabrication process. Our choice of Ge photodetector is due to its relatively lower bandgap ($E_g \sim 0.67$ eV) as it allows higher overlap of the thermal spectrum with its absorption spectrum as compared to silicon ($E_g \sim 1.1$ eV). The Ge surface is covered with a thin-film of alumina ($t_{Al_2O_3} \approx 10$ nm) to avoid electrical shorting and damage to its active region. Figure 2b shows the spectral heat flux from the metal emitter ($Cr - W - Cr$, $t_W \sim 80$ nm, $t_{Cr} \sim 5$ nm) as absorbed by a room temperature Ge photodetector ($t_{Ge} \sim 2$ μm) placed 5 micron and 50 nm apart computed using a fluctuation electrodynamics model (FED) [33–37]. The power absorbed by the photodetector increases significantly when the emitter and photodetector are separated by sub-wavelength distance. At larger gaps (far-field), the radiative power absorbed by the photodetector is much smaller due the limited thermal energy carried out by only the propagating guided modes. At sub-wavelength distances (near-field) the increase in absorbed power is due to evanescent coupling of radiation modes from the hot emitter to the cold photodetector. The details of the simulation model and the dispersion plots ($\omega - k$) for the far-field and near-field are provided in supplementary section 2. The computation results shown here are made at emitter temperature of 900 K and photodetector temperature of 293 K.

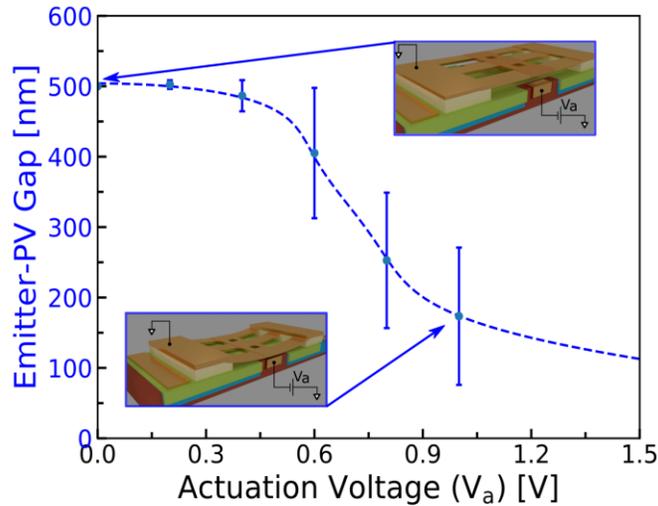

**Figure 3. NEMS characterization:** Displacement of suspended emitter as a function of applied gate-emitter potential difference ($V_a$). The broken line represents fit to a non-linear spring model (see supplementary), while the scatter points show the experimental data.

We show electrostatic control of gap, from 500 nm to 100 nm, between a suspended metallic emitter and the underlying Ge photodetector. We extract the displacement of the emitter surface by using the known initial gap ($d_0$) and measured change in capacitance caused by displacement of the bridge due to applied actuation potential ($V_a$). We confirm that the bridge is suspended and obtain the initial gap ($d_0 \sim$ 500 nm) by performing an AFM measurement between the suspended bridge and the photodetector surface (see supplementary section 3). The measured initial gap is found to be more than the thickness of deposited sacrificial layer ($t_{SiO_2} \sim$300 nm) due to bowing of the bridge under film-stress. The change in capacitance between the suspended bridge and the gate electrode as a function of gate-emitter potential ($V_a$) is measured using a vector network analyzer [38]. The details of measurement setup are provided in the supplementary section 6. Using the initial capacitance ($C_0 = \epsilon_0 \cdot A_{actuator}/d_0$) and measured change in capacitance for different gate-emitter potential, the displacement of the suspended bridge-emitter is estimated. Figure 3 shows the gap between the emitter and the photodetector surfaces as a function of applied actuation potential $V_a$ ($0\text{ V} \leq V_a \leq 1\text{V}$). One can see that the gap can be reduced from 500 nm down to ~180 nm for an applied potential of ~1V. In order to ensure that observed change in capacitance is indeed due to displacement of suspended bridge-emitter, we perform a controlled experiment using an unreleased device (sacrificial layer intact) and find negligible change in capacitance under influence of actuation potential. The raw data of the measured change in capacitance from a released and unreleased device are given in supplementary section 5. The measurement results are fitted with an electrostatic actuator model with non-linear spring response described in supplementary section 6. The multiple flexures holding the central bridge, the film stress and the partly overlapping actuation electrodes contribute to non-linear spring-like behavior allowing us to achieve gaps much smaller than those imposed by the pull-in limit of standard electrostatic actuator [39–41]. We restrict the calibration measurement of our NEMS at 1V to avoid any unwanted damage to the device prior to heat transfer measurements. The measurements shown here are performed under vacuum ($P_{\text{Chamber}} < 8 \times 10^{-5}$ Torr) to overcome the limitation of low breakdown voltage of air ($V_b \sim 0.4$ V) at extremely small distances ($d \leq d0$) in our devices.

We show an ~11x increase in generated electrical power as the distance between the suspended hot emitter and the underlying detector is reduced from ~ 500 nm to ~100 nm. We heat the thin-film emitter to high temperatures by passing electrical current ($P_{in} \approx 58$ mW) and estimate its temperature ($T_E \sim 880$ K $\pm$ 50 K) using the temperature coefficient of resistance of the film ($\alpha \approx 1.6 \times 10^{-4}$ K$^{-1}$, see methods). During the process of heating the suspended emitter we ensure that the parasitic thermal conduction into the detector is negligible ($\Delta T_D \approx 17$ K), by estimating the change in detector temperature using its IV characteristics [42,43]. Calculations for estimating the detector temperature are provided in supplementary section 4. We apply a NEMS actuation potential ($V_a$) while maintaining the emitter temperature and simultaneously measure the IV characteristics of the detector. Figure 4a shows the measured IV characteristics of the Ge detector for $0 \leq V_a \leq 1.5$. As expected, the IV characteristics shift into the power generation quadrant (4$^{\text{th}}$) with an increase in actuation potential as a result of reduction in emitter-detector distance and increase in tunneling of thermal photons [44]. In order to ensure that the measured change is indeed due to collected photons, we perform controlled experiments and measure negligible contribution of electrical leakage paths such as actuator-to-detector, emitter-to-detector and NEMS leakage current. Moreover, the measurements shown here are limited to $V_a = 1.5\ V$ as beyond this bias-point our suspended structures undergo unrecoverable damage due to excessive bending. The data for these measurements are provided in supplementary section 6 and 7. The measurements shown in Figures 4a are done under vacuum conditions (chamber pressure < $8 \times 10^{-5}$ Torr) to minimize any convective heat losses.

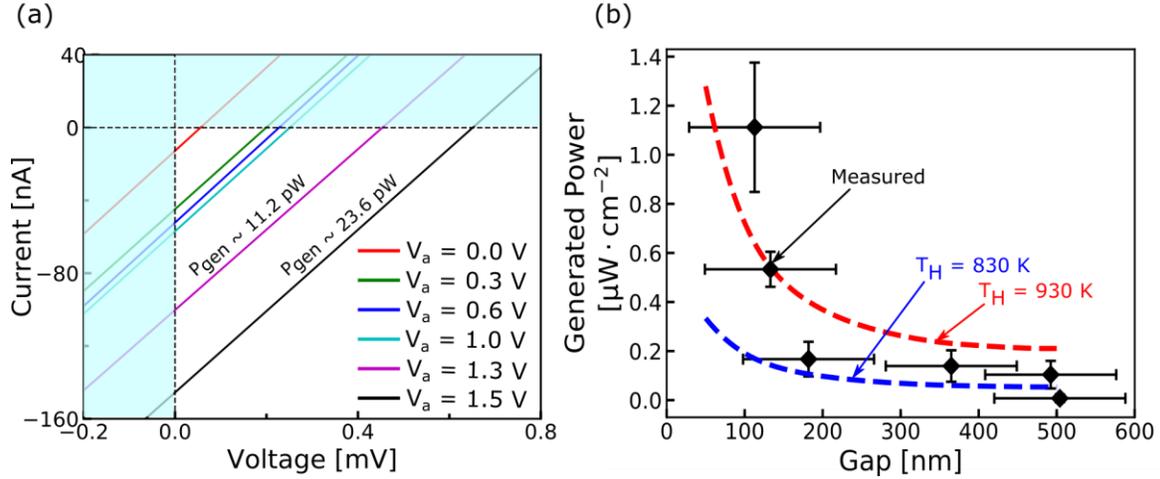

**Figure 4. Thermo-photovoltaic characterization.** (a) Measured change in IV for constant emitter temperature and varying gap achieved by applying NEMS actuation potential ($V_a$). The IV curves shift into 4th quadrant representing power generation from the near-field thermal radiation. Farther the shift, more is the power generated. (b) Measured power density from our TPV cell as a function of emitter-detector gap. The x-error bars represent the mean error in estimation of displacement while y-error bars represent the variation in power measured over two successive measurements on the same device. Also shown for is the computed power generation (broken-line) for our TPV cell with independently measured efficiency of our Ge photodetector accounted at two different emitter temperatures (confidence interval). The simulation curves also account for the bending of the suspended structure under the influence of actuation potential that is applied to reduce its distance with respect to the underlying photodetector. The simultaneously measured emitter-to-actuator leakage current is found to be <16 pA – nearly four orders smaller than the current measured from the thermal radiation.

We generate $\approx 1.25\ \mu W \cdot cm^{-2}$ from the near-field radiation between the hot Cr-W-Cr emitter at $T_E \sim 880\ K \pm 50\ K$ and the Ge photodetector at $T_D \sim 301\ K \pm 9\ K$. Figure 4b shows the generated power by our photodetector as a function of its distance from the suspended emitter. The collected power by Ge photodetector increases significantly (x11) as the gap reduces from ~500 nm to ~100 nm. The generated power ($P_{gen}$) is given by the area of the largest rectangle that fits under each IV curve - $P_{gen} = FF \times V_{oc} \times I_{sc}$; $V_{oc}$ and $I_{sc}$ are the open-circuit voltage and short-circuit current estimated from the IV characteristics of Fig. 4a, while $FF$ is the fill factor ($FF \sim 0.25$)[45]. Note that the power generated is much larger than the driving power for the capacitive NEMS structure shown here ($P_{gen}/P_{NEMS} = 10^4$; $where\ P_{NEMS} = 0.5 \cdot C_{NEMS} \cdot V_a^2$). The experimental data is seen to closely follow the theoretical prediction for the enhancement in heat transfer at small distances computed using FED. The computational data shown here takes into account the responsivity of our Ge photodetector and the bending of the emitter bridge under the influence of actuator bias $V_a$. The computation details are provided in supplementary section 2. The response of the NEMS at higher voltages ($V_a > 1\ V$) is extracted using the non-linear fit described earlier.

Our reconfigurable TPV cell can be scaled for harnessing waste heat by proper engineering of heat channels from source to emitter [46]. Leveraging the NEMS technology can allow on-demand heat recycling as well as cooling as demonstrated recently on table top experiments [47]. Our TPV cells are currently limited by the low efficiency ($\eta_D \approx 0.3 \times 10^{-4}$) of Ge photodetector due to high contact resistance and low doping concentrations of the active region (see supplementary section 7). Considering that the photodetectors at this spectral range have been widely demonstrated with efficiency as high as 10%, we expect our heat recycling efficiency could be significantly higher ($> 10 \text{ mW} \cdot \text{cm}^{-2}$) [48].

## Methods

**Fabrication process:** *Ge photodetector cell fabrication:* Germanium photodetector is fabricated on a silicon handle wafer. The epitaxial germanium (Ge) layer ($t_{Ge} \sim 1.8$ μm) is grown on a standard 6" silicon wafer using molecular beam epitaxy. After this the patterns for defining p-doping regions are transferred and then subsequently doped with boron (species: 11B$^+$, dose: $5 \times 10^{13} \text{cm}^{-2}$, power: 33 KeV). Afterwards, the p-doped regions are patterned again to define n-well regions which are then implanted with phosphorous (species: 31P$^+$, dose: $1 \times 10^{15} \text{ cm}^{-2}$, power: 50 KeV). During the implantation process the epi-Ge layer outside the patterned area is protected using ~60 nm silicon dioxide (SiO$_2$) layer deposited using plasma enhanced chemical vapor deposition (PECVD). The resulting p-n junction diode is formed in the direction perpendicular to the wafer surface. Metal contacts for the p-n diode are then deposited using e-beam evaporation technique for Ni as contact metal and Pt as the cover metal ($t_{Ni} \sim 20$ nm, $t_{Pt} \sim 45$ nm). Rapid thermal annealing is then performed to fix lattice defects from implants as well as to allow metal contact formation. The TPV cells with integrated nano-electromechanical system are fabricated on top of the Ge photodetector.

*NEMS fabrication:* In-plane actuation electrode region is carefully patterned, etched and filled up with platinum and no electrical short circuit with in-plane diode ensured. The detector and gate are covered with sacrificial layer of thin-film SiO$_2$ ($t_{SiO_2} \sim 300$ nm) using PECVD. On top of this sacrificial layer emitter patterns with their contact pads are created and tungsten (W) films ($t_W \sim 80$ nm) are sputtered (pressure: 12 mT, power: 290 W, Ar: 30 sccm) and a lift-off procedure is performed. In order to overcome the issue of adhesion and nucleation of tungsten films, we deposit thin-film of chromium ($t_{Cr} \sim 5$ nm) before and after the deposition of W. On top of patterned Cr-W-Cr emitters, we deposit amorphous silicon (a-Si) using PECVD (table temp: 350 °C, SiH$_4$ (1) +Ar (9) = 210 sccm, pressure: 1000 mT, RF power: 10 W) to provide mechanical stiffness when the bridge is suspended. The a-Si films are patterned and etched to match the underlying W emitters. The bridges are then suspended by a release step where the sacrificial SiO$_2$ is under-etched using vapor HF (SPTS Primaxx uEtch). The thin Cr layer acts as a protective layer for W-films while Al$_2$O$_3$ helps protect the Ge surface during the release step.

**Measurement of temperature coefficient:** The temperature coefficient of the W resistor is measured independently on chips before suspending the bridge-emitters. The chips are gradually heated up using a temperature-controlled sample holder and the resistance is measured at different temperatures. This is done by applying a small current (100 μA) to the W emitters to avoid any resistive self-heating effects. The slope of measured resistance for different temperatures provides the temperature coefficient of resistance ($\alpha_E \sim 0.00016 \text{ K}^{-1}$).

## Acknowledgements
The authors thank Y. Chang at National Chiao Tung University, A. Dutt at Stanford University and S.A. Miller, U.D. Dave, C.S. Joshi, M. Zadka at Columbia University, for useful discussions. The authors acknowledge the financial support from ARPA-E IDEAS program (# DE-AR0000731). The authors also acknowledge the use of facilities at Advanced Science Research Center at City University of New York, Columbia Nano-initiative, Cornell Nano-Scale Facility (NSF, # NNCI-1542081) and Cornell Center for Materials Research (NSF MRSEC, # DMR-1719875). The authors acknowledge Prof. J. Hone at Columbia University for allowing the use of vacuum probe station measurements facility at his labs.

**Author Contribution**
RS, GB, SF and ML conceived the design of the TPV. GB and RS designed the TPV cell. BZ and GB performed the heat transfer simulations and theoretical analysis. GB performed the NEMS simulations. JH provided the epitaxial germanium on silicon. GB and SR performed the fabrication of TPV cells. GB and ID performed the VNA measurements. GB performed the TPV measurements and data analysis. ID, TL and AM provided critical feedback at different stages of device design, experiments and data analysis. GB and ML prepared the manuscript. SF, BZ, RS, SR, ID, AM edited and provided feedback on the manuscript.

**Supplementary Section**
1. Device schematic
2. Theoretical analysis of thermo-photovoltaic
3. AFM data for initial height measurement
4. Estimation of diode temperature
5. Equivalent thermal circuit of the structure
6. NEMS characterization
7. Photodetector responsivity
8. Photodetector response for change in polarity of emitter heating current